# Hue Correction Scheme for Multi-Exposure Image Fusion Considering Hue Distortion in Input Images

Kouki SEO, Chihiro GO, Yuma KINOSHITA, *Student member, IEEE,* and Hitoshi KIYA, *Fellow, IEEE*
Department of Computer Science, Tokyo Metropolitan University, Tokyo, Japan

*Abstract*—We propose a novel hue-correction scheme for multi-exposure image fusion (MEF). Various MEF methods have so far been studied to generate higher-quality images. However, there are few MEF methods considering hue distortion unlike other fields of image processing, due to a lack of a reference image that has correct hue. In the proposed scheme, we generate an HDR image as a reference for hue correction, from input multi-exposure images. After that, hue distortion in an image fused by an MEF method is removed by using hue information of the HDR one, on the basis of the constant-hue plane in the RGB color space. In simulations, the proposed scheme is demonstrated to be effective to correct hue-distortion caused by conventional MEF methods. Experimental results also show that the proposed scheme can generate high-quality images, regardless of exposure conditions of input multi-exposure images.

## I. INTRODUCTION

Various high-quality imaging methods have been proposed, to obtain the wide dynamic range of real scenes [1]-[4]. Most of the methods utilize a set of differently exposed images, called "multi-exposed images," and fuse them to produce an image with high quality. These methods can be classified into two approaches. One is to tone-map a high dynamic range (HDR) image generated from multi-exposure images [1], [2]. The other is to directly fuse multi-exposure images by using a multi-exposure image fusion (MEF) method [3], [4].

The advantage of MEF compared with the former approach is that it can generate high-quality image. However, since MEF methods do not consider a non-linear camera response function (CRF), the resulting image is affected by the hue distortion in the input ones. To solve the problem, Artit et al. [5] proposed a hue-correction method for MEF. The method removes hue-distortion in an image fused by an MEF method, by using an HDR image generated form inputs as a reference for hue correction. However, this method often generates low-quality unclear images when unclear input multi-exposure images are given.

Because of the above reason, in this paper, we propose a novel hue-correction scheme for MEF. The proposed scheme generates an HDR image as a reference for hue correction from input multi-exposure images. The hue correction is performed on the basis of the constant-hue plane in the RGB color space [6], [7]. In addition, to improve the image quality when unclear input multi-exposure images are given, we use an enhance method, referred to as scene segmentation-based luminance adjustment (SSLA) [4].

Experimental results showed that resulting images of the proposed scheme has not only low hue-distortion but also high-quality.

## II. PROPOSED SCHEME

The overview of the proposed scheme is shown in Fig.1.

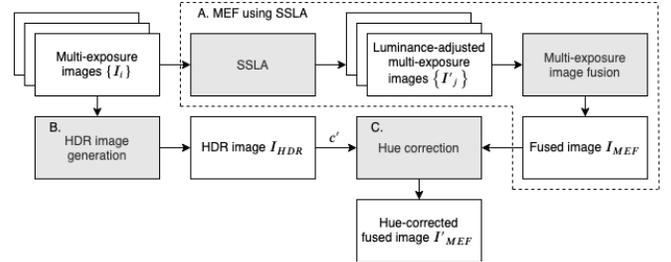

Fig. 1 Overview of proposed scheme.

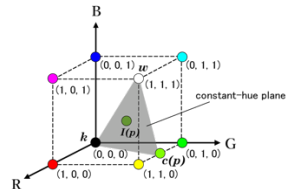

Fig. 2 Constant hue plane in RGB color space

### A. Multi-exposure image fusion using SSLA

We utilize SSLA [4] to improve the image quality of fused images by MEF, regardless of input exposure conditions. SSLA generates luminance-adjusted images from input images $I_i (1 \le i \le n, \ i \in \mathbb{N})$. The procedure of SSLA is summarized as follows:

1. Calculate the luminance $L_i$ from each input image $I_i$.
2. Obtain the enhanced luminance $L'_i$ from the luminance $L_i$ by using the dodging and burning algorithm.
3. Separate images into $m$ areas $S_j (1 \le j \le m, \ j \in \mathbb{N})$ by using a clustering algorithm based on a Gaussian mixture model (GMM).
4. Calculate the scaled luminance $\hat{L}_j$ which clearly represents an area $S_j$.
5. Tone-map the scaled luminance $\hat{L}_j$ into $\hat{L}'_j$ to avoid the saturation of luminance.
6. Calculate the adjusted pixel values of the luminance-adjusted images $I'_j$ from $\hat{L}'_j$.

After that, a fused image $I_{MEF}$ is generated as follows:
$$I_{MEF} = \mathcal{F}(I'_1, I'_2, \cdots, I'_m), \qquad (1)$$
where $\mathcal{F}(\cdot)$ is an MEF method. Here, we can use any existing MEF methods for $\mathcal{F}(\cdot)$. Mertens' MEF method [3] is used in this paper as an example.

### B. HDR image generation

To obtain reference hue information, we generate an HDR image from input multi-exposure images. The HDR image $I_{HDR}$ is generated as follows:
$$I_{HDR} = F(I_1, I_2, \cdots, I_n), \qquad (2)$$
where $F(\cdot)$ is an HDR image generation method. The method generates an HDR image by calibrating a non-linear CRF from input multi-exposure images. The calibration accuracy affects the performance of removing the non-linearity of CRF.

For this reason, we use Mitsunaga's method [1], which is one of the highest-performance HDR generation methods.

## C. Hue correction

To correct hue of image $I_{MEF}$, each pixel value $x$ in $I_{MEF}$ is represented as a linear combination on the constant hue plane [6], [7] as

$$x = a_w w + a_k k + a_c c, \quad (3)$$

where $w = (1,1,1)$, $k = (0,0,0)$ and $c$ are white, black and the maximally saturated color, as shown in Fig.2. $a_w$, $a_k$ and $a_c$ are coefficients calculated by

$$\begin{aligned} a_w &= \min(x), \\ a_k &= 1 - \max(x), \\ a_c &= \max(x) - \min(x), \end{aligned} \quad (4)$$

where $\max(\cdot)$ and $\min(\cdot)$ are functions that return the maximum and minimum elements of the pixel $x$, respectively. The maximally saturated color $c = (c_r, c_g, c_b)$ is calculated by

$$c_{r,g,b} = \frac{x_{r,g,b} - \min(x)}{\max(x) - \min(x)}, \quad (5)$$

where $x_r$, $x_g$ and $x_b$ are the R, G and B components of the pixel $x$, respectively.

Hue correction is performed by replacing the maximally saturated colors of image $I_{MEF}$ with the those of $I_{HDR}$. Let $x'$ be a pixel value of $I_{HDR}$. In accordance with Eq.(3), the pixel value $x'$ is expressed as follows

$$x' = a'_w w + a'_k k + a'_c c', \quad (6)$$

where $c'$ is a maximally saturated color calculated from the pixel value $x'$ by Eq.(5). $a'_w$, $a'_k$ and $a'_c$ are coefficients calculated by Eq.(4). Replacing $c$ with $c'$, a hue-corrected pixel value $y$ of hue-corrected image $I'_{MEF}$ is calculated as

$$y = a_w w + a_k k + a_c c'. \quad (7)$$

As a result, $y$ and $x'$ are on the same constant hue plane because they have the same maximally saturated color $c'$ that the reference HDR image has.

## III. SIMULATION

In the simulation, the performance of the proposed scheme was evaluated by comparing it with conventional ones: Mertens's MEF, Tone-mapping of estimated HDR images by Fattal's method [2], and Mertens's MEF using SSLA.

### A. Simulation condition

Hue distortion in resulting images was evaluated by using the hue difference $\Delta H$ of the CIEDE2000 [8]. The quality of resulting images generated by each method was also evaluated by using the tone mapped image quality index (TMQI) [9], which is a well-known objective quality assessment algorithm for tone-mapped images. For calculating both metrics a reference image is needed.

For this reason, 140 HDR images selected from a database [10] were used in the simulation. Input multi-exposure images with exposure values $0, \pm 2, \pm 4$ EV were generated from each HDR image.

### B. Simulation results

Figures 3 and 4 show the hue difference $\Delta H$ and TMQI scores, respectively. For the hue difference $\Delta H$, a smaller value means higher quality, and for TMQI, a larger value means higher quality. From Fig.3, it is confirmed that the proposed method had lower hue distortion than the other methods. From Fig.4, the proposed method is shown to provide high TMQI scores, regardless of exposure conditions of input multi-exposure images.

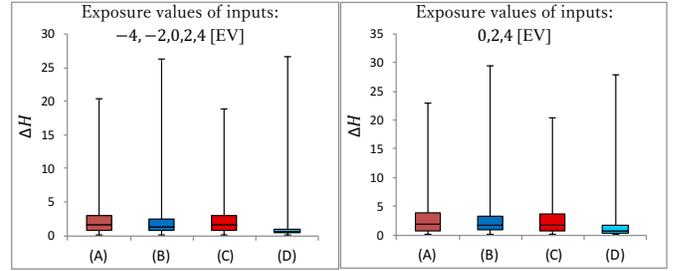

Fig. 3 $\Delta H$ scores. (A) Mertens' MEF, (B) Tone-mapping by Fattal's method, (C) Only SSLA, (D) Proposed.

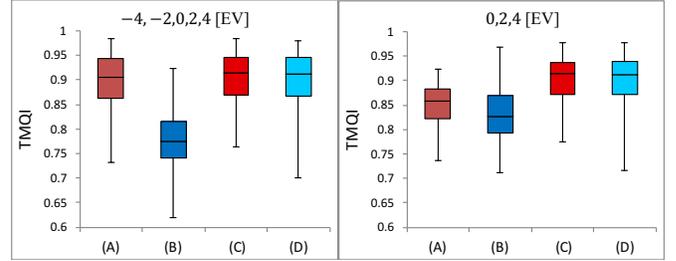

Fig. 4 TMQI scores. (A) Mertens' MEF, (B) Tone-mapping by Fattal's method, (C) Only SSLA, (D) Proposed.

Therefore, resulting images of the proposed scheme has not only low hue-distortion but also high-quality, even when unclear inputs are given.

## IV. CONCLUSION

In this paper, we proposed a novel hue-correction scheme for MEF. In the proposed scheme, a fused image is generated by a MEF method with SSLA. Then, the fused image is hue-corrected by using the hue maximally saturated colors of the HDR one generated from the same input images. From experimental results, the effectiveness of the proposed scheme was confirmed.